\documentclass{PoS}

\usepackage{amsfonts} 
\usepackage{amssymb} 
\usepackage{amsmath} 
\usepackage{subfigure}

\graphicspath{{./figs/}}

\newcommand{\MeV}{\mathop{\rm MeV}\nolimits}
\newcommand{\GeV}{\mathop{\rm GeV}\nolimits}

\title{ 
  Beyond the Standard Model B-parameters with improved staggered 
  fermions in $N_f=2+1$ QCD
}

\ShortTitle{
  BSM B-parameters with Staggered Fermions
}

\author{Taegil Bae, Yong-Chull Jang, Hwancheol Jeong,
  Jangho Kim, Jongjeong Kim, Kwangwoo Kim, Seonghee Kim, 
  Weonjong Lee, Jaehoon Leem \\
  Lattice Gauge Theory Research Center, CTP, and FPRD, \\
  Department of Physics and Astronomy,
  Seoul National University, Seoul, 151-747, South Korea \\
  E-mail: \email{wlee@snu.ac.kr}}

\author{Hyung-Jin Kim, Chulwoo Jung \\
  Physics Department, Brookhaven National Laboratory,
  Upton, NY11973, USA \\
  E-mail: \email{hjkim@bnl.gov}}

\author{Stephen R. Sharpe\\
  Physics Department, University of Washington, 
  Seattle, WA 98195-1560, USA \\
  E-mail: \email{sharpe@phys.washington.edu}}

\author{\speaker{Boram Yoon} \\
  Los Alamos National Laboratory, MS B283, P.O. Box 1663, Los Alamos, NM 87545, USA \\
  E-mail: \email{boram@lanl.gov}}

\author{SWME Collaboration}

\abstract{
  We calculate the kaon mixing B-parameters for operators arising 
  generically in theories of physics beyond the standard model.
We use HYP-smeared improved  
  staggered fermions on the $N_f = 2+1$ MILC asqtad lattices.
  Operator matching is done perturbatively at one-loop order.
  Chiral extrapolations are done using 
``golden combinations'' in which one-loop chiral logarithms
are absent.
For the combined sea-quark mass and continuum extrapolation,
we use three lattice spacings:
$a \approx 0.045,~   0.06$ and $0.09 \text{ fm}$. 
Our results have a total error of 5-6\%, which is dominated
by the systematic error from matching and continuum extrapolation.
For two of the BSM $B$-parameters, we agree with 
results obtained using domain-wall and twisted-mass dynamical fermions, 
but we disagree by $(4-5)\sigma$ for the other two.
}

\FullConference{
The 31st International Symposium on Lattice Field Theory - Lattice 2013 \\
July 29 -- August 3, 2013\\
Mainz, Germany
}

\begin{document}

\section{Introduction} 
Kaon mixing, and in particular the CP-violating component
$\epsilon_K$, can provide powerful constraints on theories of
physics beyond the standard model (BSM).
BSM theories lead to contributions to the mixing amplitude which add to
that from the standard model (SM).
Given that we now know the SM contributions to $\epsilon_K$
quite accurately, there is little room for BSM additions.

In order to determine the constraints on the parameters of BSM models,
one needs to know
the matrix elements of the local four-fermion $\Delta S=2$ operators
that arise when new, BSM heavy particles are integrated out. 
It turns out that, in addition to the operator appearing
in the short-distance component of the SM contribution (which has
a ``left-left'' Dirac structure and whose matrix element is
parametrized by $B_K$), BSM theories lead to four other operators. 
Here we present our results for the matrix elements of all five
operators, and compare with those of other 
recent calculations~\cite{Boyle:2012qb,Bertone:2012cu}.
More details of our results are given in Ref.~\cite{Bae:2013tca}.

\section{Methodology and Results \label{sec:data-anal}}
We adopt the operator basis used in Ref.~\cite{Buras:2000if},
\begin{align}
\begin{split}
 {Q}_{1} &=
 [\bar{s}^a \gamma_\mu (1-\gamma_5) d^a]
 [\bar{s}^b \gamma_\mu (1-\gamma_5) d^b]\,,   \\
 {Q}_{2} &=
 [\bar{s}^a (1-\gamma_5) d^a] [\bar{s}^b (1-\gamma_5) d^b]\,,   \\
 {Q}_{3} &=
 [\bar{s}^a \sigma_{\mu\nu}(1-\gamma_5) d^a]
 [\bar{s}^b \sigma_{\mu\nu} (1-\gamma_5) d^b]\,,   \\
 {Q}_{4} &=
 [\bar{s}^a (1-\gamma_5) d^a] [\bar{s}^b (1+\gamma_5) d^b]\,,  \\
 {Q}_{5} &=
 [\bar{s}^a \gamma_\mu (1-\gamma_5) d^a]
  [\bar{s}^b \gamma_\mu (1+\gamma_5) d^b]
\,,
\end{split}
\label{eq:operators}
\end{align}
where $a$, $b$ are color indices and 
$\sigma_{\mu\nu} =  [\gamma_\mu, \gamma_\nu]/2$.
$Q_1$ is the SM operator while 
$Q_{2-5}$ are the BSM operators.
Matrix elements of the latter are parametrized as
\begin{align}
\label{eq:def-B_i}
\begin{split}
  B_i(\mu) = 
\frac{\langle \overline{K}_0 \vert Q_i (\mu) \vert K_0 \rangle} 
{N_i \langle \overline{K}_0 \vert\overline{s}\gamma_5 d(\mu)\vert 0 \rangle 
    \langle 0 \vert \bar{s} \gamma_5 d (\mu) \vert K_0 \rangle}
\\
  (N_2,\ N_3,\ N_4,\ N_5) = ( 5/3, \ 4, \ -2, \ 4/3 )
\,.
\end{split}
\end{align}
The basis (\ref{eq:operators}) is that in which the
the two-loop anomalous dimensions 
(which we use for renormalization group running) are known~\cite{Buras:2000if}.
It differs from the ``SUSY'' basis of Ref.~\cite{Gabbiani:1996hi} 
(which has been used in previous lattice
calculations of the BSM matrix elements),
although the two bases are simply related, as discussed below.

\begin{table}[tbp]
\caption{MILC lattices used in this calculation.
  Here ``ens'' is the number of gauge configurations,
  ``meas'' is the number of measurements per configuration,
  and ID identifies the ensemble.
  \label{tab:milc-lat}}
\begin{center}
\begin{tabular}{c  c  c  c  l }
\hline\hline
$a$ (fm) & $am_\ell~/~am_s$ & \ \ size & ens $\times$ meas  & ID \\
\hline
0.09  & 0.0062~/~0.031 & $28^3 \times 96$  & $995 \times 9$ & F1 \\
0.09  & 0.0093~/~0.031 & $28^3 \times 96$  & $949 \times 9$ & F2 \\
0.09  & 0.0031~/~0.031 & $40^3 \times 96$  & $959 \times 9$ & F3 \\
0.09  & 0.0124~/~0.031 & $28^3 \times 96$  & $1995\times 9$ & F4 \\
0.09  & 0.00465~/~0.031 & $32^3 \times 96$  & $651\times 9$ & F5 \\
\hline
0.06  & 0.0036~/~0.018 & $48^3 \times 144$ & $749 \times 9$ & S1 \\
0.06  & 0.0072~/~0.018 & $48^3 \times 144$ & $593 \times 9$ & S2 \\
0.06  & 0.0025~/~0.018 & $56^3 \times 144$ & $799 \times 9$ & S3 \\
0.06  & 0.0054~/~0.018 & $48^3 \times 144$ & $582 \times 9$ & S4 \\
\hline
0.045 & 0.0028~/~0.014 & $64^3 \times 192$ & $747 \times 1$ & U1 \\
\hline\hline
\end{tabular}
\end{center}
\end{table}

Table~\ref{tab:milc-lat} shows the 
lattices used in this work.
They are generated using $N_f=2+1$ flavors of 
asqtad staggered quarks~\cite{Bazavov:2009bb}.
For valence quarks, we use HYP-smeared staggered 
fermions~\cite{Hasenfratz:2001hp}.

The calculation of the BSM B-parameters follows closely the methodology
used in our $B_K$ calculation~\cite{Bae:2010ki, Bae:2011ff}.
The lattice operators are matched, at one-loop order,
to continuum operators defined in the renormalization
scheme of Ref.~\cite{Buras:2000if}.
This scheme differs slightly from that
used in the continuum-lattice matching calculation of
Ref.~\cite{Kim:2011pz}, requiring an additional (continuum) matching 
factor which we have calculated.
The matching on each lattice is done at a
renormalization scale $\mu=1/a$, 
with results subsequently run to a common final scale using the
continuum two-loop anomalous dimensions.

We use ten different valence quark masses,
$  am_{x,y} = am_s \times \frac{n}{10}$
(with $n=1-10$),
where $m_x$ and $m_y$ are the valence $d$ and $s$ quark masses, respectively,
while $m_s$ (given in Table~\ref{tab:milc-lat})
lies close to the physical strange quark mass.
Thus our valence pion masses run down almost to 200 MeV.
Results are extrapolated to physical light-quark masses using SU(2)
staggered chiral perturbation theory (SChPT),
taking the lightest four quark masses for 
$m_x$ and the heaviest three for $m_y$.
Thus we remain in the regime, $m_x \ll m_y \sim m_s$,
where we expect SU(2) ChPT to be applicable.

\begin{figure}[btp]
\begin{center}
\includegraphics[width=20pc]{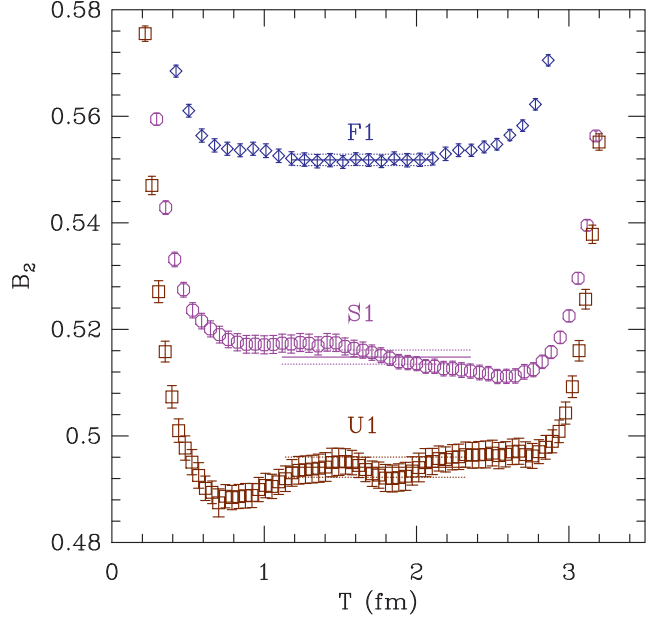}
\end{center}
\caption{$B_2(\mu=1/a)$ as a function of $T$, the
distance between the left-hand kaon source and the operator.
(Blue) diamonds, (purple) octagons and (brown) squares are from 
the F1, S1, and U1 ensembles, respectively, with 
$(m_x, ~m_y) = (m_s/10, ~m_s)$.
\label{fig:b_2}}
\end{figure}

As an example of the quality of our data, we show,
in Fig.~\ref{fig:b_2}, the plateaus that
we find for $B_2$ on the three lattice spacings.
This is for our most kaon-like choice of valence quark masses.
We use U(1) noise sources to create kaons with a fixed
separation, and place the lattice operators between them.
From plots such as these, as well as those for kaon correlators,
we determine how far from the sources we must work to avoid
excited-state contamination. We then fit to a constant.

%

Generalizing a proposal of Ref.~\cite{Becirevic:2004qd}, 
Ref.~\cite{Bailey:2012wb} suggested that chiral extrapolations 
of BSM $B$-parameters would
be simpler if one uses the following ``golden combinations'':
\begin{align}
G_{23} = \frac{B_2}{B_3}, \quad
G_{45} = \frac{B_4}{B_5}, \quad
G_{24} = B_2 \cdot B_4, \quad
G_{21} = \frac{B_2}{B_K}.
\label{eq:golden}
\end{align}
This is because the leading chiral logarithms,
which appear at next-to-leading order (NLO) in ChPT,
cancel in these quantities (if one uses SU(2) ChPT).
This observation is particularly important for staggered fermions,
since chiral logarithms introduce taste-breaking effects which
normally have to be corrected for.
Hence, we perform the chiral and continuum extrapolations using
the golden combinations, as well as $B_K$, 
and obtain our final results by inverting Eq.~(\ref{eq:golden}).
%
%
We obtain consistent results by
directly extrapolating the $B$-parameters themselves
using the SU(2) SChPT forms of Ref.~\cite{Bailey:2012wb}, 
following the analysis method previously followed for 
$B_K$~\cite{Bae:2010ki, Bae:2011ff}.

To extrapolate in $m_x$, we fit the golden combinations to
\begin{equation}
G_i(\text{X-fit}) = 
 c_1 + c_2 X + c_3 X^2 + c_4 X^2 \ln^2 X 
 + c_5 X^2 \ln X + c_6 X^3 \,, 
\end{equation}
where $X \equiv X_P / \Lambda_\chi^2$ with $X_P = M_{\pi:x\bar
  x}^2$, and $\Lambda_\chi = 1\GeV$, and generic NNLO continuum
chiral logarithms are included.
We call this the X-fit.
To fit our four data points to this form, we constrain
the coefficients $c_{4-6}$ with Bayesian priors: $c_i = 0 \pm 1$.
Systematic errors in X-fits 
are estimated by doubling the widths of these priors,
and by comparing to the results of the eigenmode shift 
method~\cite{Jang:2011fp}.
The X-fits for $B_K$ do involve NLO chiral logarithms, and we
follow the same procedure as in Refs.~\cite{Bae:2010ki, Bae:2011ff}.

\begin{figure}[tbp]
\centering
\subfigure[X-fit]{
\label{subfig:X-fit}
\includegraphics[width=17pc]{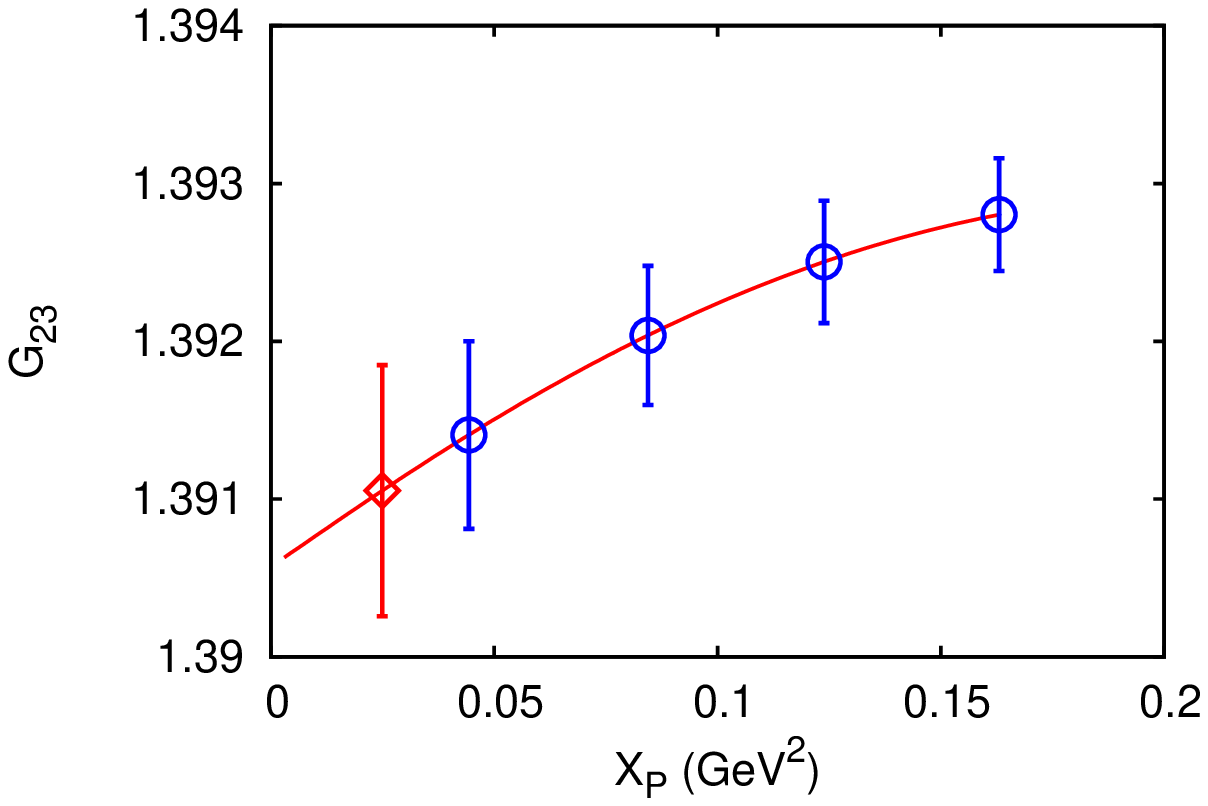}
}
\subfigure[Y-fit]{
\label{subfig:Y-fit}
\includegraphics[width=17pc]{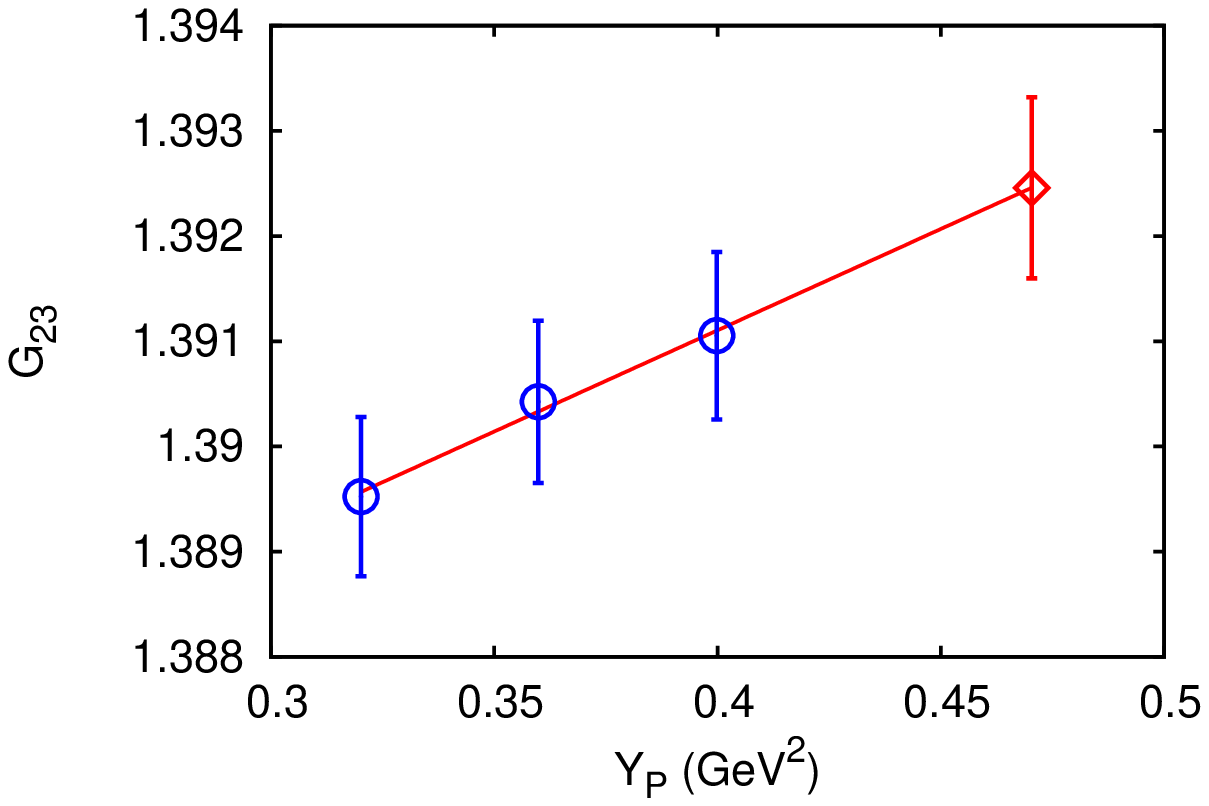}
}
\caption{ (a) X-fit and (b) Y-fit of $G_{23}(\mu=1/a)$ on F1 ensemble.
  In the X-fit, $m_y$ is fixed at $am_y = 0.03$.
  The red diamond represents the extrapolated physical point in both plots.
  \label{fig:chiral-fit}}
\end{figure}

The extrapolation in the valence strange mass $m_y$
is done using a linear function of $Y_P = M_{\pi:y\bar y}^2$, 
with a quadratic fit used to estimate a systematic error.
We call this the Y-fit.
In Fig.~\ref{fig:chiral-fit}, we show X- and Y-fits
for $G_{23}$, illustrating the mild
dependence on the valence quark masses.

\begin{figure}[tbp]
\centering
\subfigure[\ $B_K$]{
\label{subfig:X-fit}
\includegraphics[width=17pc]{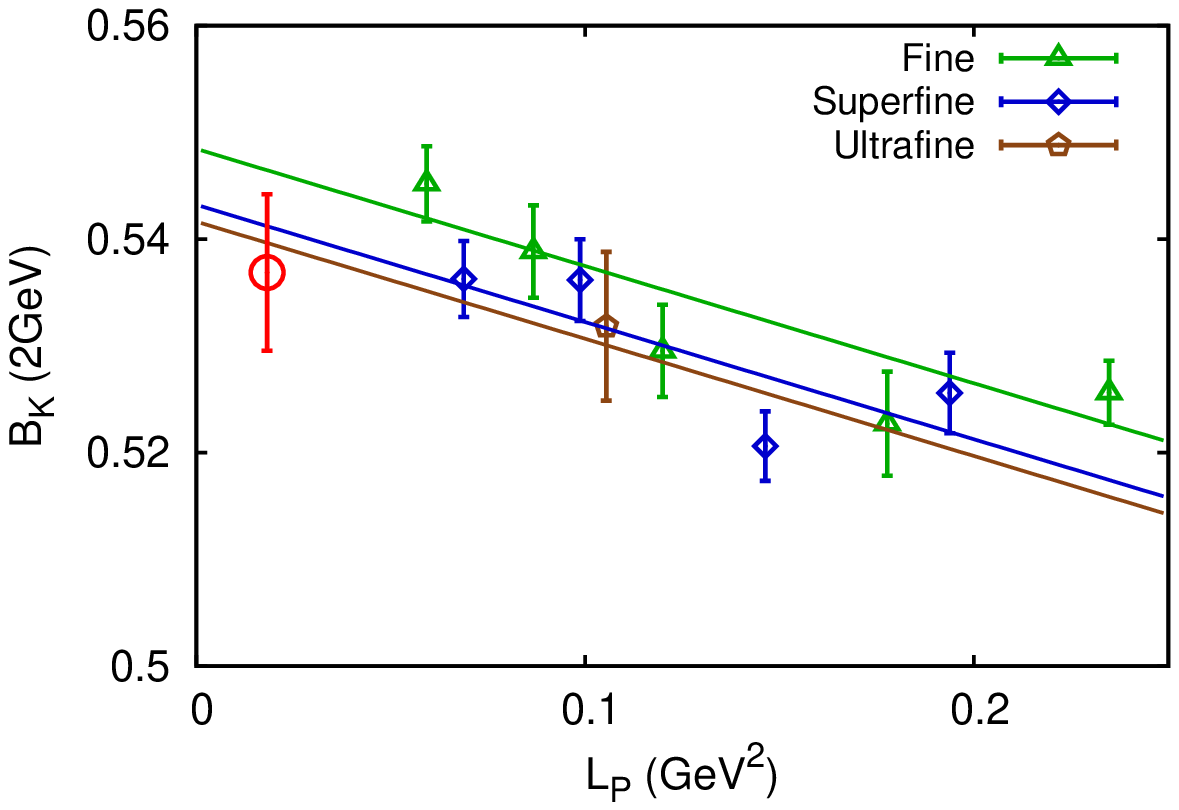}
}
\subfigure[\ $G_{23}$]{
\label{subfig:Y-fit}
\includegraphics[width=17pc]{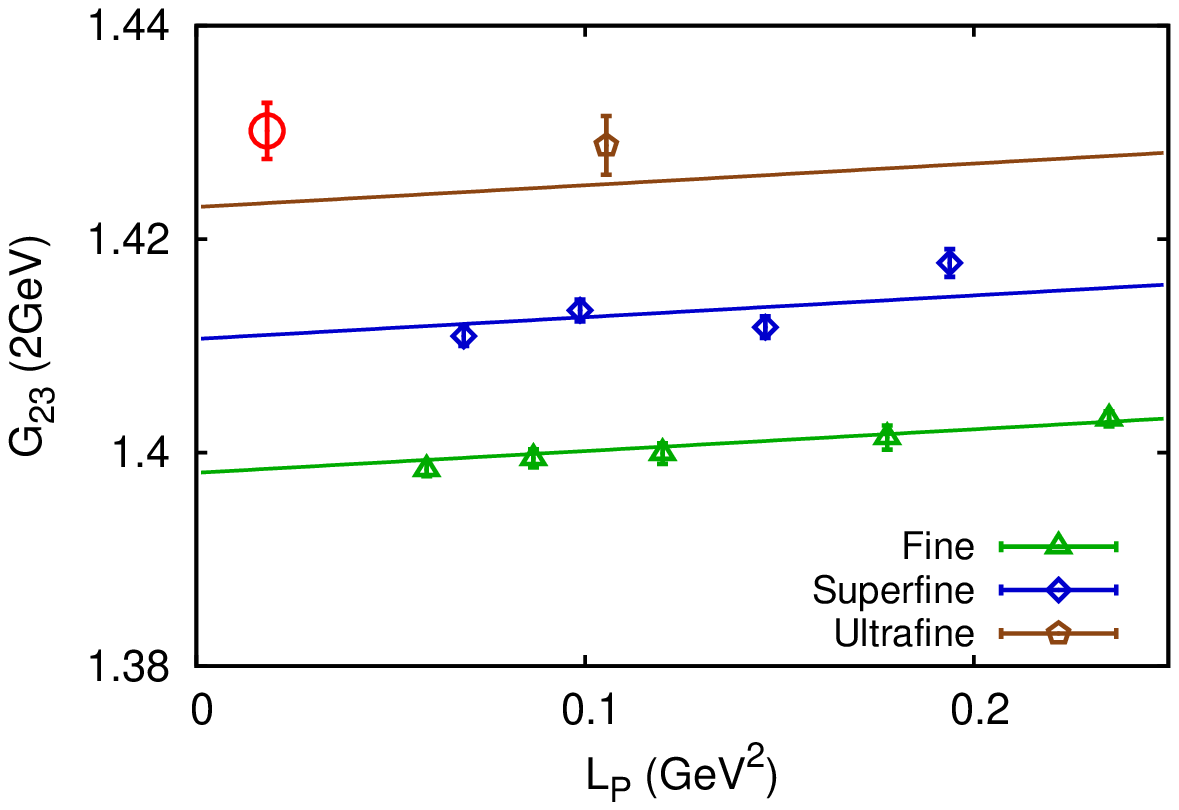}
}
\caption{Examples of simultaneous sea-quark mass 
and continuum extrapolations.
Red circles denote extrapolated results.
  \label{fig:scaling}}
\end{figure}

After these valence chiral extrapolations,
we have, for each ensemble, results for the 
the $G_i$ (and $B_K$) at scale $\mu=1/a$. 
We next evolve these to a common scale,
either $2\GeV$ or $3\GeV$, using the two-loop anomalous dimensions
of Ref.~\cite{Buras:2000if}.
These results can then be extrapolated to the continuum and to physical
sea-quark masses (with the dependence on the latter being analytic at NLO).
We do a simultaneous extrapolation using
\begin{equation}
f_1=d_1 + d_2 (a \Lambda_Q)^2 + d_3 {L_P}/{\Lambda_\chi^2}
+ d_4 {S_P}/{\Lambda_\chi^2}\,,
\label{eq:fitf1}
\end{equation}
where $L_P$ and $S_P$ are squared masses of taste-$\xi_5$ pions composed of
two light sea quarks and two strange sea quarks, respectively.
Discretization errors are scaled with $\Lambda_Q = 300\MeV$,
so that for a typical discretization error one
would have $d_2\sim\mathcal{O}(1)$.
However, we find that $d_2 \sim 2 - 7$ for the golden combinations, 
indicating enhanced lattice artifacts.
The artifacts in $B_K$ are much smaller.
These results are illustrated in Fig.~\ref{fig:scaling}.
%
The extrapolation fits have
$\chi^2/\text{dof}$ in the range $1.6 - 2.7$.

To estimate the systematic uncertainty in the continuum 
extrapolation, we compare the results of fitting to
Eq.~\eqref{eq:fitf1} with those obtained using
the following fitting function, which includes higher order terms 
in $a^2$ and $\alpha_s$ [which here means
$\alpha_s^{\overline{\rm MS}}(1/a)$]:
\begin{equation}
f_2 = f_1 + d_5 (a\Lambda_Q)^2 \alpha_s + d_6 \alpha_s^2
+ d_7 (a\Lambda_Q)^4\,.
\label{eq:fitf2}
\end{equation}
The difference, $\Delta f_{12}=|f_1-f_2|$, is about $5\%$, and is 
comparable with our estimate of
the systematic error coming from the one-loop matching, 
which we estimate by $\alpha_s^2(\text{U1}) = 4.4\%$.
Since Eq.~\eqref{eq:fitf2} includes terms of $\alpha_s^2$, it also captures the
systematic error from using one-loop perturbative matching.
Hence, we quote the larger of $\Delta f_{12}$ and 
$\alpha_s^2(\text{U1})$ in our final error budget for the
matching/continuum extrapolation error. 
%

%
%
\begin{table}[tbp]
\caption{Results for the $B_i$ at $\mu=2\GeV$ and $3\GeV$.}
  \label{tab:b-par}
\begin{center}
\begin{tabular}{ l c c c c }
\hline\hline
      & $\mu=2\GeV$ & $\mu=3\GeV$ \\
\hline
$B_K$ & 0.537 (7)(24)  & 0.519 (7)(23) \\
$B_2$ & 0.620 (4)(31)  & 0.549 (3)(28) \\
$B_3$ & 0.433 (3)(19)  & 0.390 (2)(17) \\
$B_4$ & 1.081 (6)(48)  & 1.033 (6)(46) \\
$B_5$ & 0.853 (6)(49)  & 0.855 (6)(43) \\
\hline\hline
\end{tabular}
\end{center}
\end{table}
\begin{table}[tbp]
\caption{Error budget (in percent) for the $B_i(2\;\GeV)$.}
  \label{tab:err-budget}
\begin{center}
\begin{tabular}{ l  c c c c c }
\hline\hline
\quad source of error & $B_K$ & $B_2$ & $B_3$ & $B_4$ & $B_5$ \\
\hline
\quad statistics      & 1.37 & 0.64 & 0.63 & 0.60 & 0.66 \\
$\left\{ \begin{array}{l} \text{matching} \\ \text{cont-extrap.} \end{array} \right\}$
                      & 4.40 & 4.95 & 4.40 & 4.40 & 5.69 \\
\quad X-fit (F1)      & 0.10 & 0.10 & 0.10 & 0.12 & 0.12 \\
\quad Y-fit (F1)      & 0.62 & 0.12 & 0.19 & 0.22 & 0.16 \\
\quad finite volume   & 0.50 & 0.50 & 0.50 & 0.50 & 0.50 \\
\quad $r_1=0.3117(22)\;$fm
                      & 0.34 & 0.18 & 0.17 & 0.05 & 0.02 \\
\quad $f_\pi=132$ vs. $124\MeV$ (F1)
                      & 0.46 & 0.46 & 0.46 & 0.46 & 0.46 \\
\hline\hline
\end{tabular}
\end{center}
\end{table}

In Table~\ref{tab:b-par}, we show our final results for the B-parameters
evaluated at $\mu=2\GeV$ and $3\GeV$.
Total errors are $5-6\%$, and are dominated the systematic errors,
as can be seen from the error budget given in Table~\ref{tab:err-budget}.
Further details concerning error estimates are explained in
Ref.~\cite{Bae:2013tca}.
Overall, we see that the matching/continuum extrapolation dominates.

\section{Comparisons and Outlook}
As noted above, 
there are two previous calculations of the BSM $B$-parameters 
using dynamical  quarks: 
one using $N_f=2+1$ flavors of domain-wall fermions
at a single lattice spacing~\cite{Boyle:2012qb},
the other using $N_f=2$ flavors of twisted-mass fermions at
three lattice spacings~\cite{Bertone:2012cu}.
The results from these two calculations are consistent.
We pick those of Ref.~\cite{Boyle:2012qb} for a detailed comparison
with our results, since then both calculations
use the same number of dynamical flavors.

We begin by noting that the results for $B_K$ in all three calculations
are consistent (within the small errors).
For the BSM B-parameters, Ref.~\cite{Boyle:2012qb} finds,
at $\mu=3\GeV$, that
$B_{2-5}^\text{SUSY} = 0.43(5)$, $0.75(9)$, $0.69(7)$ and $0.47(6)$, 
respectively.
These are calculated in the SUSY basis, 
in which $B_3^\textrm{SUSY} = (5 B_2-3 B_3)/2$,
while the other three $B$-parameters are the same.
Our results convert (at $\mu=3\GeV$) to $B_3^\textrm{SUSY} = 0.79(3)$.
Using this, and Table~\ref{tab:b-par}, we see that,
while $B_2$ and $B_3$ are consistent, $B_4$ and $B_5$ are not.
Our results are 1.5 and 1.8 times larger, respectively,
a difference of $4-5\sigma$.

At the present time, we do not know the origin of this difference.
We have ruled out the simplest possibilities, such as using incorrect
continuum anomalous dimensions, by various cross checks with results
in the literature.
It thus seems that systematic errors in one or both
calculations are being underestimated, the most likely culprit
being the estimate of the truncation error in matching.
This enters at the two-loop level in both calculations, since both are
ultimately connected to a continuum scheme by one-loop matching.

We think it is important to resolve this disagreement, not only
because the matrix elements are wanted for phenomenology, but also
because it may impact the reliability of other calculations.
One way to proceed is for all calculations to use a
common (S)MOM scheme
and non-perturbative renormalization and running.
This avoids the matching to the continuum.  
We are working in this direction.
Another idea is for the other calculations to do the chiral
extrapolations using the golden combinations described above.

\section*{Acknowledgments}
We thank Claude Bernard for providing unpublished information.  
W.~Lee is supported by the Creative Research Initiatives program
(2013-003454) of the NRF grant funded by the Korean government (MSIP).
C.~Jung and S.~Sharpe are supported in part by the US DOE through
contract DE-AC02-98CH10886 and grant DE-FG02-96ER40956, respectively.
Computations for this work were carried out in part on the QCDOC
computer of the USQCD Collaboration, funded by the Office of Science
of the US DOE.  W.~Lee acknowledges support from the KISTI
supercomputing center through the strategic support program
[No. KSC-2012-G3-08].
%

\bibliographystyle{JHEP}
\bibliography{ref}

\end{document}